\newif\ifdraft
\newif\iffull
\newif\ifcomment
\newif\iflatexdiff
\newif\ifbibtex
\newif\ifpreprint
\newif\ifbanner
\newif\ifsupp
\bibtextrue      
\preprinttrue    

\def\dvers{v1}

\ifpreprint

\fi
\def\dtitle{Event-shape and multiplicity dependence of freeze-out radii\\in pp collisions at $\mathbf{\sqrt{{\textit s}}}=7$ TeV}
\def\stitle{Event-shape and multiplicity dependence of freeze-out radii in pp}
\ifpreprint
\documentclass[ALICE,manyauthors,12pt]{cernphprep}
\usepackage[comma,square,numbers,sort&compress]{natbib}
\usepackage{hyperref}
\else
\documentclass[final,3p,12pt]{elsarticle}
\biboptions{comma,square,numbers,sort&compress}
\usepackage{hyperref}
\fi
\usepackage{graphicx}  
\usepackage{dcolumn}   
\usepackage{bm}        
\usepackage{amssymb}   
\usepackage{amsfonts}
\usepackage{graphics}
\usepackage{grffile}   
\usepackage{epsfig}
\usepackage{units}
\usepackage[usenames]{color}
\usepackage[normalem]{ulem} 
\usepackage{color}
\usepackage[utf8]{inputenc}
\usepackage[T1]{fontenc}
\usepackage{subfigure}

\iflatexdiff
\RequirePackage{color}\definecolor{RED}{rgb}{1,0,0}\definecolor{BLUE}{rgb}{0,0,1}

\fi

\ifpreprint
\else
 
\fi

\newcommand{\fc}           {\ensuremath{f_{\rm c}}}
\newcommand{\ST}           {\ensuremath{S_{\rm T}}}
\newcommand{\kT}           {\ensuremath{k_{\rm T}}}

\newcommand{\Nchr}         {\ensuremath{N_{\rm ch}^{\rm rec}}}

\newcommand{\dNdeta}       {\ensuremath{{\rm d}N_{\rm ch}/{\rm d}\eta}}
\newcommand{\avdNdeta}     {\ensuremath{\left<\dNdeta\right>}}

\newcommand{\pt}           {\ensuremath{p_{\mathrm{T}}}{ }}

\newcommand{\abs}[1]       {\ensuremath{\left|#1\right|}}

\newcommand{\red}[1]       {\textcolor{red}{#1}}

\newcommand{\warn}[1]      {{\small\textbf{\red{(!}\footnote{\textbf{\red{(!)}}~#1}\red{)}}}\marginpar{\textbf{\red{---}}}}

\newcommand{\com}[1]       {}

\iflatexdiff

\renewcommand{\xout}[1]    {\textcolor{red}{\sout{#1}}}
 
\newcommand{\old}[1]       {{\textcolor{red}{\sout{#1}}}}

\else

\renewcommand{\xout}[1]    {}
\newcommand{\old}[1]       {\relax}

\fi

\ifdraft
\usepackage{lineno}
\linenumbers
\modulolinenumbers[1]
\usepackage{fancyhdr}
\ifbanner
\fi

\else
\renewcommand{\warn}[1]{}

\fi
\begin{document}
\newlength{\figlen}
\setlength{\figlen}{\linewidth}
\ifpreprint
\setlength{\figlen}{0.95\linewidth}
\begin{titlepage}
\PHyear{2018} 
\PHnumber{332}                        
\PHdate{10 December}                     
\title{\dtitle}
\ShortTitle{\stitle}
\Collaboration{ALICE Collaboration%
         \thanks{See Appendix~\ref{app:collab} for the list of collaboration members}}
\ShortAuthor{ALICE Collaboration} 
\ifdraft
\begin{center}
\ifbanner
 \today\\ \color{red}DRAFT \dvers\ \hspace{0.3cm} \$Revision: 1777 $\color{white}:$\$\color{black}\vspace{0.3cm}
\else
 \today\\ \color{red}DRAFT \dvers\ \color{black}\vspace{0.3cm}
\fi
\end{center}
\fi
\else
\begin{frontmatter}
\title{\dtitle}
\iffull
\input{Alice_Authorlist_2014-Mar-21-PLB}
\else
\ifdraft
\author{ALICE Collaboration \\ \vspace{0.3cm} 
\today\\ \color{red}DRAFT \dvers\ \hspace{0.3cm} (\$Revision: 1777 $\color{white}:$\$)\color{black}}
\else
\author{ALICE Collaboration}
\fi
\fi
\fi
\begin{abstract}
\
Two-particle correlations in high-energy collision experiments enable the extraction of particle source radii by using the Bose-Einstein enhancement of pion production at low relative momentum $q\propto 1/R$.
It was previously observed that in $\rm{p}\rm{p}$ collisions at $\sqrt{s}=7$ TeV the average pair transverse momentum~\kT~range of such analyses is limited due
to large background correlations which were attributed to mini-jet phenomena.
To investigate this further, an event-shape dependent analysis of Bose-Einstein correlations for pion pairs is performed in this work.
By categorizing the events by their transverse
sphericity \ST~into spherical $(S_\textrm{T}>0.7)$ and jet-like
$(S_\textrm{T}<0.3)$ events a method was developed that allows for the determination of source radii for much larger values
of \kT~for the first time. 
Spherical events demonstrate little or no background correlations while jet-like events are dominated by them. This observation agrees with the hypothesis of a mini-jet origin of the non-femtoscopic background correlations and gives new insight into the physics interpretation of the \kT~dependence of the radii. 
The emission source size in spherical events shows a substantially diminished \kT~dependence, while jet-like events show indications of a negative trend with respect to \kT~in the highest multiplicity events. Regarding the emission source shape, the correlation functions for both event sphericity classes show good agreement with an exponential shape, rather than a Gaussian one.

\ifdraft 
\ifpreprint
\end{abstract}
\end{titlepage}
\else
\end{abstract}
\end{frontmatter}
\newpage
\fi
\fi
\ifdraft
\thispagestyle{fancyplain}
\else
\end{abstract}
\ifpreprint
\end{titlepage}
\else
\end{frontmatter}
\fi
\fi
\setcounter{page}{2}


\section{Introduction}
\label{sec:intro}
Bose-Einstein correlations for pairs of identical bosons with low relative momentum are essential tools for understanding
particle production in ultra-relativistic collision
experiments~\cite{Heinz:1999rw,Wiedemann:1999qn,Lisa:2005dd}. 
They allow one to extract the dimensions of the freeze-out stage of the reaction, usually known as the ``source radii''. A linear dependence of the volume defined by such radii on the charged-particle multiplicity 
produced in the event was observed in pp, p--Pb, and Pb--Pb collisions at the Large Hadron Collider
(LHC)~\cite{Khachatryan:2010un,Aamodt:2010jj,Khachatryan:2011hi,Aamodt:2011kd,Adam:2015pya,Abelev:2014pja,Aad:2015sja}.
Interestingly, pp collisions at the LHC have reached multiplicities
similar to the ones obtained in peripheral p--Pb and Pb--Pb
collisions, thus allowing for a direct comparison of their source
radii~\cite{Aamodt:2011kd,Adam:2015pya,Abelev:2014pja}.  

In proton-proton collisions, correlations of non-identical pions $(\pi^+\pi^-)$ show significant deviations from unity particularly in low multiplicity events and at high pair transverse momentum $k_\textrm{T}=\frac{1}{2} \left| {\bf p}_1+{\bf p}_2\right|_{\rm T}$, which are attributed to resonance decays and fragmentation of mini-jets from low momentum-transfer scatterings~\cite{Aamodt:2011kd,Adam:2015pya}. These correlations are a dominant background to the Bose-Einstein correlations between pairs of identical pions $(\pi^+\pi^++\pi^-\pi^-)$ and make interferometry analyses at the LHC challenging for \kT~greater than about 0.6 GeV/$c$. However, the three-pion cumulant approach significantly suppresses the mini-jet related backgrounds in pp and p-Pb collisions~\cite{Abelev:2014pja}, though it can suffer from statistical limitations.

In this paper, we introduce a new way of reducing the mini-jet background in pp collisions by selecting events based on their transverse sphericity~\cite{Abelev:2012sk,Banfi:2004nk}, an observable that is sensitive to particle collimation and as such can differentiate between jet-like (hard) and spherical (soft) event topologies.
Such a differential measurement based on sphericity, \kT, and multiplicity, will offer new insights into (mini-jet induced) background correlations and offer invaluable information needed to improve the accuracy of event generators, such as PYTHIA~\cite{Sjostrand:2006za,Skands:2010ak}. As will be shown, this advancement can significantly extend the \kT~reach of interferometry analyses in general.

This paper presents two-particle correlation functions (CFs) as a function of the pair relative three-momentum 
$q=\sqrt{\left( p_1- p_2 \right)^i \left( p_1- p_2 \right)_i}$ and the source radius parameter for different intervals of $dN_{\textrm{ch}}/d\eta$ and \kT~for jet-like and spherical event topologies.
In particular, the \kT~dependence of the radii in spherical events is investigated since small background correlations in these events allows for the study of possible signs of collectivity in pp collisions.

\section{Experimental setup and data selection}
\label{sec:setup}
Approximately $5\times 10^8$ pp collisions at $\sqrt{s}=7$ TeV were analyzed, which were recorded by the ALICE experiment at the LHC~\cite{Aamodt:2008zz,Abelev:2014ffa} during the 2010 running period.

The main detectors used for this analysis are: the Inner Tracking System (ITS)~\cite{Aamodt:2010aa}, the Time Projection Chamber (TPC)~\cite{Alme:2010ke}, the Time-Of-Flight detector (TOF)~\cite{Cortese:2002kf}, and the V0~\cite{Abbas:2013taa}.
The ITS is a six-layer cylindrical silicon detector used for precise vertex and track reconstruction close to the interaction point. It provides full azimuthal coverage and spans the pseudorapidity range $\left|\eta\right|<0.9$. 
The TPC is the main tracking detector in ALICE and measures the specific ionization energy loss of particles in the TPC gas for particle identification (PID). It covers the whole azimuth and provides a radial coverage of 159 possible space points for tracks. It fully covers the $\left|\eta\right|<0.9$ range while extending out to $\left|\eta\right|<1.5$ with a smaller number of potentially reconstructed space points.
The TOF uses multigap resistive plate chambers to measure particle arrival time and thus particle velocity. 
It extends the PID capabilities to the intermediate $p_\textrm{T}$ range where the pion, proton, and kaon energy loss signals are similar in the TPC.
The V0 detectors are used for triggering on collision events. They are composed of two small-angle scintillator arrays, located at 340 cm and $-90$ cm from the nominal interaction point along the beam line.
The minimum-bias trigger, which is used in this analysis, requires at least one hit in the V0 or either of the two first layers of the ITS in coincidence with two beam bunches crossing in the ALICE interaction region, which is measured by a beam-pickup system. An offline event selection is applied to reject beam-halo induced events and beam-gas collisions.

Accepted events have their primary vertex reconstructed within $\pm 8$ cm from the center of the detector along the beam line in order to ensure uniform tracking performance. Charged particle tracks are reconstructed with the ITS and TPC detectors, requiring that each TPC track segment is reconstructed from at least 70 out of the 159 possible space points. 
To guarantee that mainly primary particles are selected it is required that the track has its Distance of Closest Approach (DCA) to the primary vertex smaller than $(0.0182+0.35 \cdot p^{-1.01}_{\textrm{T}})$ cm in the transverse plane, with $p_\textrm{T}$ in GeV/\textit{c}, and 0.2 cm in the longitudinal direction.
Tracks with a kink topology in the TPC, indicating weak decays of charged kaons, are rejected.
Two-track effects such as merging and splitting are minimized using pion pair selection criteria as described in~\cite{Aamodt:2011kd} and are known to be negligible in this \pt range for $q$ greater than about 50 MeV/\textit{c}, which is much less than the expected width of the Bose-Einstein correlation peak.

\section{Analysis technique}
\label{sec:analysis}

The interferometry analysis was performed using pions with pseudorapidity $\abs{\eta}<1.2$ and transverse momentum $0.13<\pt<4.0$ GeV/\textit{c}. 
Pions were identified by their specific ionization energy loss in the TPC as well as the measured pion arrival time in the TOF detector. The PID selection criteria are the same as described in~\cite{Aamodt:2011kd}. They are optimized to maximize efficiency while producing a high-purity of the pion sample of about $99$\% for $p_{\rm T}<2.5$ GeV/\textit{c}. The \pt resolution is about 1\% or better in the relevant \pt range.

Transverse sphericity is calculated by using all charged tracks with $\abs{\eta}<0.8$ and $\pt>0.5$ GeV/\textit{c}.  In order to avoid a bias from the boost along the beam axis~\cite{Banfi:2004nk}, the event shape is calculated only in the transverse plane.
The transverse sphericity matrix $S_{\textrm{XY}}$ is defined as
\begin{equation}
S_{\textrm{XY}}=\frac{1}{\sum_{i} p_{\textrm{T}}^{i}} \sum_{i} \frac{1}{p_{\textrm{T}}^{i}}
\begin{pmatrix} 
(p_x^{i})^2 & p_x^{i} \cdot p_y^{i} \\
p_x^{i} \cdot p_y^{i} & (p_y^{i})^2
\end{pmatrix},
\label{equation_ST}
\end{equation}
where $i$ runs over all charged particle tracks in the event.
By using the transverse sphericity matrix eigenvalues $\lambda_1$ and $\lambda_2$~\cite{Abelev:2012sk}, the transverse sphericity is computed as
\begin{equation}S_{\textrm{T}}=\frac{2\cdot \textrm{min}(\lambda_1\textrm{, } \lambda_2) }{\lambda_1+\lambda_2}.
\end{equation}

It is a bounded scalar observable that is sensitive to the event shape, and in particular particle collimation.
An event with only one hard scattering will in general produce a jet-like distribution that yields low sphericity while multiple soft scatterings are expected to yield high sphericity events. Events with several independent hard scatterings can also yield higher sphericity as each (mini-)jet axis is oriented randomly. Transverse sphericity is known to be correlated with the number of hard parton-parton interactions in an event~\cite{Abelev:2012sk}. In this work jet-like events with $S_\textrm{T}<0.3$ and spherical events with $S_\textrm{T}>0.7$ are analyzed, which comprise 18\% and 28\% of the total minimum-bias data set, respectively.

The resolution in \ST~due to finite track reconstruction efficiency is found to be better than 0.1 based on Monte-Carlo (MC) simulations with the PYTHIA 6.4 Perugia-0 tune~\cite{Sjostrand:2006za,Skands:2010ak}, for both low- and high-sphericity events in any multiplicity interval. No significant effect from the finite track momentum resolution was observed in simulations.

\begin{table}[ht]
\begin{center}
  \begin{tabular}{ c  c  c }
    \hline    \hline~
    $\Nchr$ & $\avdNdeta_{\ST<0.3}$ & $\avdNdeta_{\ST>0.7}$\\ \hline
    [1,13] & $4.3\pm 2.0$ & $5.3\pm 2.2$ \\ [0pt]
    [14,21] & $9.6\pm 2.0$ & $10.5\pm 2.1$ \\ [0pt]
    [22,30] & $13.9\pm 2.2$ & $14.9\pm 2.3$ \\ [0pt]
    [31,54] & $18.7\pm 3.2$ & $20.4\pm 3.2$ \\[0pt]
    \hline\hline
  \end{tabular}
\end{center}
\caption{Intervals of $\Nchr$ and corresponding mid-rapidity $\avdNdeta$ with $\abs{\eta}<1.2$ and  $0.13<\pt<4.0$ GeV/\textit{c} for both sphericity ranges.}
\label{table:1}
\end{table}

The multiplicity estimator $N_{\rm ch}^{\rm rec}$ is defined as the number of reconstructed charged-particle tracks that enter the interferometry analysis. 
The intervals and their corresponding corrected $\avdNdeta$ for both sphericity selections are shown in Table~\ref{table:1}. 

\subsection{Two-pion correlation function analysis}
\label{sec:CFs}
The one-dimensional femtoscopic analysis presented in this paper was performed
using the invariant momentum difference \textit{q}, which corresponds to
the magnitude of the relative three-momentum in the pair rest frame (PRF).
The measured CFs are defined as the ratio of the $q$ distributions 
for same-event $(A)$ and mixed-event $(B)$ pion
pairs times a normalization factor $(\xi)$~\cite{Lisa:2005dd},
\begin{equation}
C(q;N_{\rm ch}^{\rm rec},S_{\rm T}) =
\xi(N_{\rm ch}^{\rm rec},S_{\rm T}) \cdot
\frac{A(q;N_{\rm ch}^{\rm rec},S_{\rm T})}{B(q;N_{\rm ch}^{\rm
    rec},S_{\rm T})}.
\label{eq:Cdef}
\end{equation}
For the event mixing, pools of eight events of similar multiplicity and
sphericity are formed. In addition, it is also required that events in a mixed event pool have their vertex positions within 2 cm from each other in the beam direction. The mixed event distributions were then made by pairing up pions from different events in a mixed event pool. Identical selection criteria are applied to the same-event and mixed-event pion pairs, and both $A(q)$ and $B(q)$ are constructed in the same \ST~interval.

 Both distributions are normalized in the range $0.7 <q < 0.8$ GeV/\textit{c}, which is well outside the relevant quantum statistical (QS) domain ($q\propto 1/R\approx 0.3$ GeV/\textit{c}) and below the onset of the high-\textit{q} rise associated with energy and momentum conservation.

\begin{figure}[ht]
  \centering
  \includegraphics[width=0.7\textwidth]{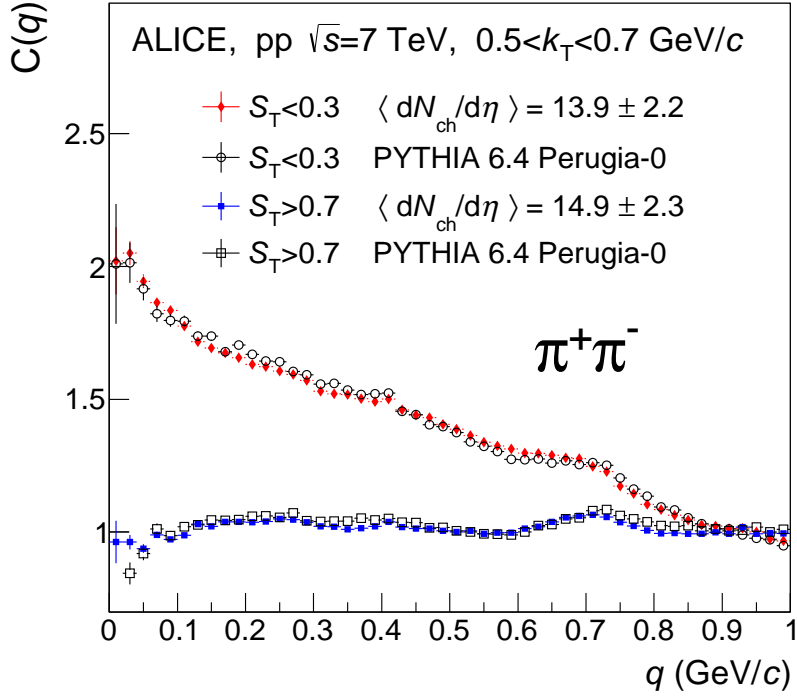}
  \caption{Opposite-sign pion pair correlation functions in data and PYTHIA simulations for high and low sphericity intervals. The error bars represent statistical uncertainties.}
  \label{plot:1}
\end{figure}

Figure~\ref{plot:1} shows good agreement between the measured correlation functions for opposite-sign pion pairs in pp collisions at $\sqrt{s}=7$~TeV and PYTHIA simulations, which include the ALICE detector response, for spherical and jet-like events at similar multiplicity. 
Opposite-sign CFs do not include Bose-Einstein correlations but do include backgrounds, such as those induced by mini-jets, which are also found in same-sign pair analyses~\cite{Aamodt:2011kd}. 
They also show features due to two-body decays like $\textrm{K}^0_{\rm S}$ and $\rho\rightarrow \pi^+ \pi^-$ at about $412$ and $723$ MeV/\textit{c} in $q$ respectively, and a wide three-body decay peak from $\omega\rightarrow \pi^+ \pi^-\pi^0$.

The $C(q)$ in spherical events are relatively flat at unity, while the low \ST~CFs exhibit a very pronounced slope. This finding supports previous assumptions about the mini-jet origin of background correlations in interferometry analyses~\cite{Aamodt:2011kd} and demonstrates that PYTHIA describes two-pion correlations well in the absence of Bose-Einstein correlations. 

\begin{figure}[ht]
  \centering
  \includegraphics[width=0.7\textwidth]{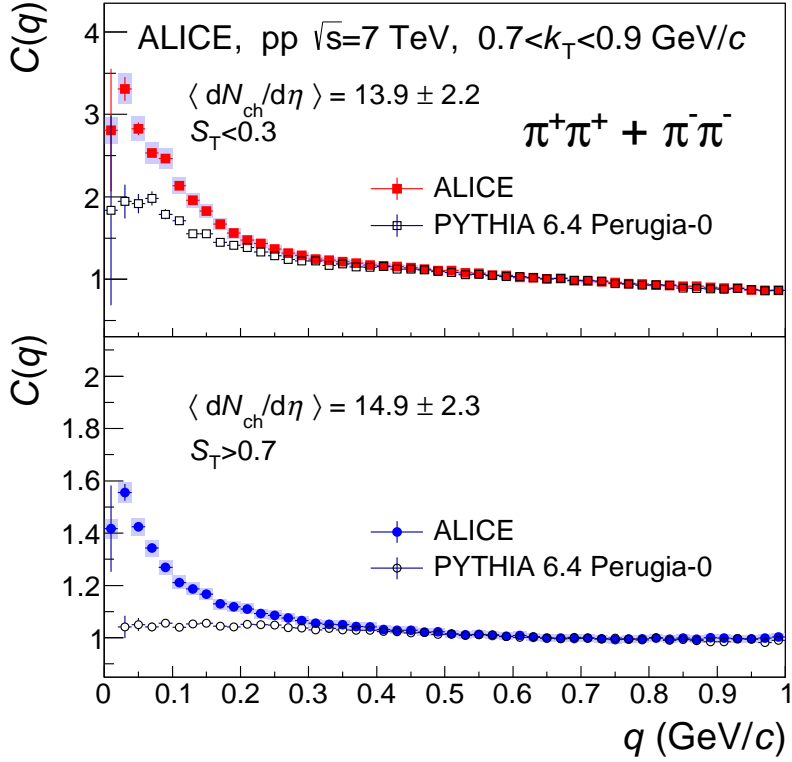}
  \caption{Comparison of same-sign correlation functions for data and PYTHIA simulations for both sphericity intervals. The error bars represent statistical uncertainties and the shaded boxes are systematic uncertainties. Note that the vertical axes have different scales in the two panels.}
\label{data_vs_MC_pic}
\end{figure}

Figure \ref{data_vs_MC_pic} shows a comparison of CFs for same-sign pion pairs from data and PYTHIA simulations for the two sphericity intervals at similar reconstructed multiplicity. The MC includes neither quantum-statistical correlations nor final-state interactions (FSI).

Similar to the opposite-sign CF at $S_T>0.7$ shown in
Fig.~\ref{plot:1}, the spherical-event same-sign CF is rather flat
outside the QS correlation region~($q<0.5$
  GeV/\textit{c}). This indicates that the background is small
  in spherical events for same-sign pairs.The shape of the same-sign 
$C(q)$ in spherical events is compatible with the expectation
from Bose-Einstein correlations. There are no novel features like peaks or depressions and the correlation function does not extend outside the theoretically predicted values $1 \le C(q) \le 2$. On the other hand, for jet-like events, the CF exhibits a pronounced slope over the full $q$ range, indicating the presence of background. The CF shape is well described by PYTHIA outside the QS correlation region. Moreover, it is observed that the large-$q$ correlation increases with \kT~and decreases with multiplicity, which is consistent with previous findings in ~\cite{Aamodt:2011kd}. These results suggest that the primary source of background correlations in two-pion femtoscopic analyses is related to semi-hard scattering processes which predominantly populate
jet-like event topologies.

To extract the source size from the measured correlation functions, background correlations are corrected with simulated CFs $(C_{\rm{MC}}(q))$. This procedure assumes that the signal and background factorize and was used in a similar way by other experiments~\cite{Khachatryan:2010un,Aad:2015sja}. The method also resembles \cite{Sirunyan:2017ies} where the background signal is not extracted from MC simulations but fitted in the measured opposite-sign correlations and then removed out via the fitting procedure in the same-sign analysis. In the case of our analysis this approach showed to be unstable.

The simulations used for the corrections include the ALICE detector response and are analyzed exactly the same as the data.
 The corrected correlation function $\widetilde{C}(q)$ can then be expressed as
\begin{equation}
  \centering
  \widetilde{C}(q) = \frac{C_{\rm data}(q)}{C_{\rm MC}(q)} = \frac{C_{\rm BE+FSI}(q) \cdot C_{\rm ES}(q) }{C^{\rm MC}_{\rm ES}(q)},
  \label{eq:C2corrections}
\end{equation}
where $C_{\rm BE+FSI}$ and $C_{\rm ES}$ are contributions coming from Bose-Einstein correlations with FSI effects and event shape dependent backgrounds, respectively. 
The corrected CFs, $\widetilde{C}(q)$, are obtained separately for spherical and jet-like events. 
The femtoscopic correlations are then determined using
\begin{equation}
  \centering
  \widetilde{C}(q)=\left[ (1-\fc^2) + \fc^2 K(q) C^{\rm QS}(q)\right],
  \label{eq:C2QSextraction}
\end{equation}
as in~\cite{Sinyukov:1998fc}, where $\fc^2$ is the pair fraction from
the core of the particle-emission source~\cite{Csorgo:1998tn}, $K(q)$
is the FSI correlation and $C^{\rm QS}(q)$ is the extracted QS
correlation.  
The $K(q)$ factor is well known and calculated using the two-pion FSI
wave functions \cite{Lednicky:2005tb} that include Coulomb and strong
interactions. 
The values of $\fc^2$ are estimated from
EPOS-LHC MC model~\cite{Pierog:2013ria}, which is known to reproduce a variety
of LHC measurements.  
The deviation of $\fc^2$ from unity quantifies the degree of dilution
caused by pions from long-lived resonances and weak decays. This effect is suppressed by the track selection used. In previous
measurements~\cite{Abelev:2014pja}, $\fc^2$ showed little dependence
on \kT\ up to $0.7$ GeV/\textit{c} and calculations using EPOS-LHC
agreed with this observation at even larger values of \kT.  
At all \kT, the final value of $\fc^2$ depended on the track selection,
leading to larger values of $\fc^2$ in cases where a tighter DCA
selection was used~\cite{Abelev:2013pqa}. 
In this analysis, the $\fc^2$ is fixed to 0.85 which corresponds to its \kT-averaged
value. Finally, a fit is applied to the extracted $C^{\rm QS}(q)$ correlations in
order to determine the femtoscopic radii.

\begin{figure}[hbt]
  \centering
  \includegraphics[width=0.65\textwidth]{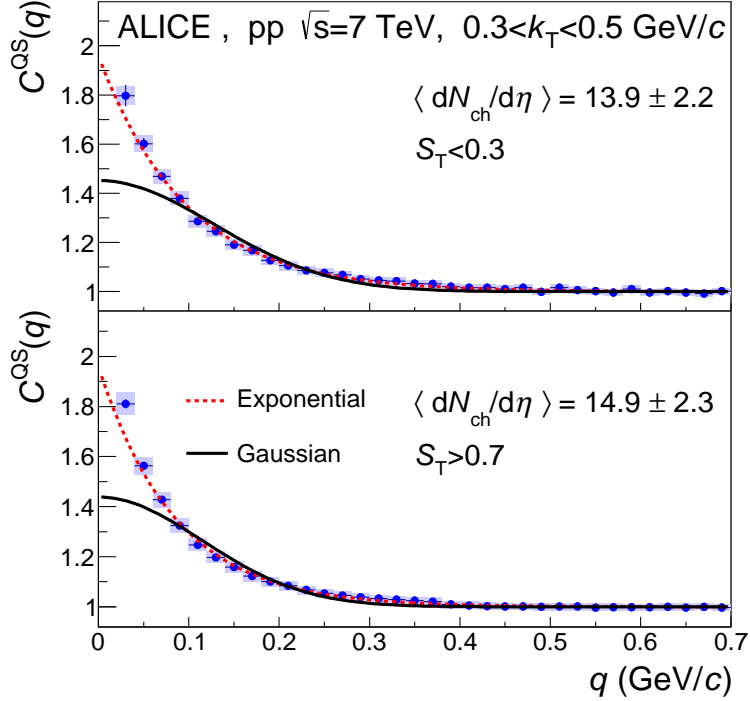}
  \caption{Comparison of exponential and Gaussian fit results for $C^{QS}(q)$ functions in spherical and jet-like events.}
  \label{fig:C2QSfit}
\end{figure}

In past analyses~\cite{Bialas:2000yi}, it was
observed that a Gaussian form of $C^{\rm QS}(q)$ does not describe
the observed one-dimensional CFs over the full $q$ range. Previous
analyses at the LHC showed that this remains the case at much higher
collision energy ~\cite{Khachatryan:2010un,Aamodt:2010jj}.
Hence, a Levy fit is performed, employing a free parameter $\alpha$ in the exponent, where $\alpha=2$ corresponds to a Gaussian distribution. Figure \ref{fig:C2QSfit} shows the fits of $C^{\rm QS}(q)$ with Gaussian and exponential functional forms, and it is observed that the 1D CFs are better described by $\alpha=1$ corresponding to an exponential distribution. Therefore, femtoscopic radii are extracted assuming an exponential shape:
\begin{equation}
  \centering
  C^{\rm QS}(q)=1+\lambda\cdot e^{\left(-R_{inv}\cdot q\right)}.
  \label{eq:C2QS}
\end{equation}
In the fitting procedure, $\lambda$ was first treated as a fit parameter and showed negligible \kT~dependence.
The mean value of $\lambda$ was then obtained for each sphericity interval and fixed for the final fitting. Since the expected dilution from long-lived resonance decays and weak decays is explicitly removed in Eq.~\ref{eq:C2QSextraction}, the $\lambda$ parameter is expected to be consistent with unity in the case of fully chaotic emission and exponential 1D correlation functions.
At $k_T<0.7$ GeV/\textit{c}, the mean value of $\lambda$ is 0.97 for spherical and 1.0 for jet-like events, with small deviations consistent with statistical fluctuation in each individual measurement.

To extract radii the correlation functions are fit with the same functional form for each \kT~interval while fixing $\lambda$ to the aforementioned mean values. No additional normalization factors are used in fitting $C^{\rm QS}(q)$.

A previous sphericity-integrated measurement~\cite{Aad:2015sja}, using a significantly different fitting procedure and thus having an alternative interpretation of the $\lambda$ parameter to the one used in this paper, observed a stronger \kT\ dependence of both the radii and $\lambda$ parameters.
Considering the different parameterizations used in \cite{Abelev:2013pqa}, model calculations and previous measurements~\cite{Abelev:2013pqa}, it is unlikely that there exists such a significant change in the fraction of long-lived emitters or pion coherence to explain such a strong decrease of $\lambda$ with \kT.

\subsection{Systematic uncertainty}
\label{sec:sys}

Analyses were performed using data from different data-taking periods with varying experimental conditions (e.g.
detector operating conditions, polarity of the magnetic field in the
apparatus, etc.) and showed negligible differences in the observed radii. 
Similarly, a 100\% increase in the number of events that are used in the mixing procedure also produced negligible differences in the measured radii.
Separate analyses for positively and negatively charged pions, which, at LHC energies, are expected to give identical results, gave less than 0.2\% differences in measured radii.

Regarding more substantial uncertainties in the measured radii a distinction was made between point-by-point uncorrelated and correlated sources of systematic uncertainty. It was observed that there are two significant sources of uncorrelated systematic uncertainty, 
the first one being variations in the tracking procedure, and the second one being variations in the CF fit range. The uncorrelated uncertainty due to the tracking procedure is evaluated by using an alternative track selection for the analysis, in which only the TPC is used to reconstruct tracks as in~\cite{Aamodt:2011kd}, and is estimated to be up to $10$\% on the measured radii.
Concerning the uncertainty due to the fit range selection, in this analysis $q<0.7$ GeV/\textit{c} was used as the default fit
range while $q<0.4$ GeV/\textit{c} and 
$q<1.0$ GeV/\textit{c} were the variations. A difference in radii of up
to 5\% is observed in this case with the smaller fit range always having a larger
influence on the change of radii, as is expected.

In this analysis, the correlated uncertainties in the measured radii are shown to be larger than the uncorrelated ones and are estimated by varying $f_c^2$, sphericity ranges and Monte Carlo generators. 
A variation in $f_c^2$ of $\pm0.05$ produced a $5$--$10$\% uncertainty,
with the largest deviation being observed in the highest $N_{\rm ch}$
and \kT\ bin. The variation of the sphericity range, which was varied by $\pm0.05$, contributed up to 10\% in the systematic uncertainty. The leading source of correlated uncertainty, which shows a difference in the radii of up to 15\% depending on the \kT~and multiplicity bin, is the choice of Monte Carlo generators for the background correlations. In this analysis we used PYTHIA as the default and PHOJET~\cite{Engel:1994vs} as the variation. As is expected, the spherical event results are less effected by this variation than the jet-like event ones.  
\section{Results}
\label{sec:results}

Figure 4 shows the measured $R_{\textrm{inv}}$ as a function of pair \kT~for spherical and jet-like events in different multiplicity intervals. In comparison to previously
observed 1D radii~\cite{Aamodt:2011kd}, the multiplicity- and \kT-dependence observed in the event shape dependent analysis show both similarities and significant differences depending on the sphericity selection. In spherical events the dependence of the radii on \kT~is well
described by a constant. As is expected, the fitted constant radius
increases with multiplicity from $1.971 \pm 0.006 \textrm{ (stat.)} \pm 0.106\textrm{ (sys.)}$~fm in the lowest multiplicity bin to $2.410 \pm 0.007 \textrm{ (stat.)} \pm 0.050\textrm{ (sys.)}$~fm at the highest multiplicity. On the other hand, similar to previous sphericity-integrated results, the
jet-like radius dependence on \kT~is not well described by a 0th-order
polynomial ($\chi^{2}/N_{\textrm{dof}}$ is larger than 10), however a 1st-order polynomial does manage to describe the data better. The constant of this fit is $1.97 \pm 0.01 \textrm{ (stat.)} \pm 0.11\textrm{ (sys.)}$~fm at lowest multiplicity and increases to $2.40 \pm 0.05 \textrm{ (stat.)} \pm 0.05\textrm{ (sys.)}$~fm at
highest multiplicity. The slope parameter is observed to be negative except for the lowest multiplicity bin where it is consistent with zero. For higher multiplicities the difference from a zero slope is observed as a 3$\sigma$, 7.3$\sigma$ and 4.1$\sigma$ effect consecutively.

\begin{figure}[t]

  \centering
  \includegraphics[width=0.75\textwidth]{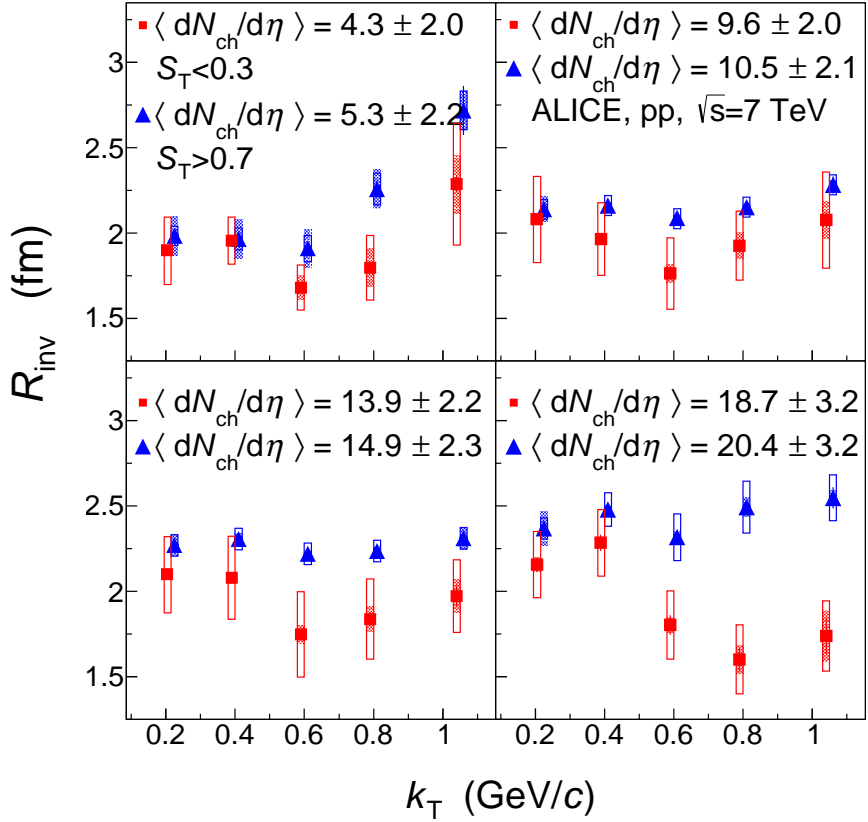}
  \caption{Measured 1D source radii as a function of pair \kT~for
    spherical and jet-like events in pp collisions at $\sqrt{s}=7$
    TeV. Points are shifted horizontally for
    clarity. The vertical error bars represent the statistical uncertainties of the measurement, while the shaded and open boxes represent the uncorrelated and correlated systematic uncertainties, respectively.} 
\label{radii}
\end{figure}

In spherical events the overall size of the system grows with
increasing event multiplicity. Such a trend has been observed in all previous measurements in proton-proton and heavy-ion collisions. Spherical events are expected to contain multiple scatterings, so a growth of system size
with multiplicity is naturally expected.  On the other hand, jet-like
events are typically dominated by a single interaction with a
large momentum transfer and subsequent fragmentation into
(mini-)jets where most of particle production occurs. 

Our data suggest that the hard scattering process has a
non-trivial space-time structure which is consistent with some
theoretical predictions~\cite{Paic:2005cx}. It is not possible to provide a more direct interpretation of the results, because the modeling of the space-time
structure of the fragmentation process is largely neglected in current
Monte-Carlo codes such as PYTHIA and PHOJET.

The \kT~dependence of $R_{inv}$ in jet-like events is much less explored
theoretically, due to the reasons given above. Therefore, our
measurements present a pioneering insight into the space-time
characteristics of fragmentation. It is observed that for higher multiplicity ranges the \kT~dependence has a non-zero slope, for example $-0.52 \pm 0.05 \textrm{ (stat.)} \pm 0.05\textrm{ (sys.)}$ fm/(GeV/\textit{c}) for the second highest multiplicity selection, obtained from a linear fit to the radii, taking into account all sources of systematic uncertainties. These results raise an interesting question on the origin of the \kT~dependence in minimum bias 1D measurements by ALICE~\cite{Aamodt:2011kd}, especially if considered together with the lack of such a decrease in spherical events. It is confirmed that by averaging the radii from spherical, jet-like and intermediate events ($0.3 < S_{\rm T} < 0.7$, not shown in this work) it is possible to reproduce the dependence observed in the sphericity-integrated analysis.  The fact that in a differential measurement the slope is most prominent in jet-like events suggests that it is this category of events that might be the dominant source of the non-zero \kT~slope in the minimum-bias data set as well. 

The development of a Monte-Carlo code that fully incorporates the
space-time structure of proton-proton collisions and the fragmentation process
would be highly desirable, a candidate for such an event generator would be  EPOS~\cite{Pierog:2013ria}. The predictions of such a model, as a function of event sphericity, should then be carefully compared with existing and future data on pion Bose-Einstein correlations. Further improvements in the modeling of experimental transverse sphericity distributions is also desirable.

In a three-dimensional analysis of femtoscopic radii in the
Longitudinally Co-Moving System (LCMS), which is the reference system
where pair longitudinal momentum vanishes, with the Pratt-Bertsch
decomposition of the momentum difference, where the ``long'' direction
is along the beam, ``out'' is along pair transverse momentum and
``side'' is perpendicular to the other two, a decreasing trend with
\kT~is interpreted as evidence of hydrodynamic
collectivity~\cite{Adam:2015vna}. This reasoning has also been applied
to small systems~\cite{Pierog:2013ria}, therefore this dependence
is a critical test of the interpretation of spherical and jet-like
events presented above. 
However, it was shown that the \kT~dependence in the PRF is additionally 
influenced by the boost factor between the LCMS and the PRF~\cite{Shapoval:2013jca,Kisiel:2014upa}. Therefore, in this work we are not able to draw conclusions about collectivity in pp collisions based on the \kT~dependence observed
in Fig.~\ref{radii}.

\section{Summary}
\label{sec:summary}
In summary, the measured pp collisions at $\sqrt{s}=7$ TeV have been classified into sub-samples with high (``spherical'') and low (``jet-like'') sphericity and source radii have been extracted for both.
An exponential fit function, as opposed to a Gaussian, was shown to better describe the observed 1D correlation functions for both sphericity ranges. 
A significant suppression of non-femtoscopic correlations was observed
in ``spherical'' events, effectively doubling the \kT~range of
femtoscopic analyses for this sample. 
Substantial background correlations remained in the ``jet-like''
sample, which is consistent with the hypothesis that the main source
of these correlations is mini-jets. PYTHIA and PHOJET describe this background, making an effective background removal procedure feasible. As a consequence, radii for jet-like
events have been extracted for the first time. They tend to be smaller in comparison to source radii of spherical events, which may be a consequence of lower average multiplicities, and show a decrease with \kT, resembling the trend observed also in minimum-bias analyses. The extracted radii in spherical events show an increase in the system size with multiplicity, in agreement with previous measurements and model expectations. They are also observed to have a flat trend in \kT, which differs from the \kT~dependence that was previously observed in 1D minimum-bias analyses~\cite{Khachatryan:2010un,Aamodt:2010jj,Khachatryan:2011hi,Aamodt:2011kd,Adam:2015pya,Abelev:2014pja,Aad:2015sja}. This suggests that the observed slope in minimum-bias events could be arising from the lower part of the transverse sphericity spectrum in pp collisions.
This is novel and unique information on the  space-time characteristics of the fragmentation process. Future investigations will require more advanced modeling of the space-time properties of particle production in Monte-Carlo codes, and a comparison of such theoretical predictions to a three-dimensional pion femtoscopy measurement performed differentially in transverse sphericity. 

\newenvironment{acknowledgement}{\relax}{\relax}
\begin{acknowledgement}
\section*{Acknowledgements}

The ALICE Collaboration would like to thank all its engineers and technicians for their invaluable contributions to the construction of the experiment and the CERN accelerator teams for the outstanding performance of the LHC complex.
The ALICE Collaboration gratefully acknowledges the resources and support provided by all Grid centres and the Worldwide LHC Computing Grid (WLCG) collaboration.
The ALICE Collaboration acknowledges the following funding agencies for their support in building and running the ALICE detector:
A. I. Alikhanyan National Science Laboratory (Yerevan Physics Institute) Foundation (ANSL), State Committee of Science and World Federation of Scientists (WFS), Armenia;
Austrian Academy of Sciences and Nationalstiftung f\"{u}r Forschung, Technologie und Entwicklung, Austria;
Ministry of Communications and High Technologies, National Nuclear Research Center, Azerbaijan;
Conselho Nacional de Desenvolvimento Cient\'{\i}fico e Tecnol\'{o}gico (CNPq), Universidade Federal do Rio Grande do Sul (UFRGS), Financiadora de Estudos e Projetos (Finep) and Funda\c{c}\~{a}o de Amparo \`{a} Pesquisa do Estado de S\~{a}o Paulo (FAPESP), Brazil;
Ministry of Science \& Technology of China (MSTC), National Natural Science Foundation of China (NSFC) and Ministry of Education of China (MOEC) , China;
Croatian Science Foundation and Ministry of Science and Education, Croatia;
Centro de Aplicaciones Tecnol\'{o}gicas y Desarrollo Nuclear (CEADEN), Cubaenerg\'{\i}a, Cuba;
Ministry of Education, Youth and Sports of the Czech Republic, Czech Republic;
The Danish Council for Independent Research | Natural Sciences, the Carlsberg Foundation and Danish National Research Foundation (DNRF), Denmark;
Helsinki Institute of Physics (HIP), Finland;
Commissariat \`{a} l'Energie Atomique (CEA) and Institut National de Physique Nucl\'{e}aire et de Physique des Particules (IN2P3) and Centre National de la Recherche Scientifique (CNRS), France;
Bundesministerium f\"{u}r Bildung, Wissenschaft, Forschung und Technologie (BMBF) and GSI Helmholtzzentrum f\"{u}r Schwerionenforschung GmbH, Germany;
General Secretariat for Research and Technology, Ministry of Education, Research and Religions, Greece;
National Research, Development and Innovation Office, Hungary;
Department of Atomic Energy Government of India (DAE), Department of Science and Technology, Government of India (DST), University Grants Commission, Government of India (UGC) and Council of Scientific and Industrial Research (CSIR), India;
Indonesian Institute of Science, Indonesia;
Centro Fermi - Museo Storico della Fisica e Centro Studi e Ricerche Enrico Fermi and Istituto Nazionale di Fisica Nucleare (INFN), Italy;
Institute for Innovative Science and Technology , Nagasaki Institute of Applied Science (IIST), Japan Society for the Promotion of Science (JSPS) KAKENHI and Japanese Ministry of Education, Culture, Sports, Science and Technology (MEXT), Japan;
Consejo Nacional de Ciencia (CONACYT) y Tecnolog\'{i}a, through Fondo de Cooperaci\'{o}n Internacional en Ciencia y Tecnolog\'{i}a (FONCICYT) and Direcci\'{o}n General de Asuntos del Personal Academico (DGAPA), Mexico;
Nederlandse Organisatie voor Wetenschappelijk Onderzoek (NWO), Netherlands;
The Research Council of Norway, Norway;
Commission on Science and Technology for Sustainable Development in the South (COMSATS), Pakistan;
Pontificia Universidad Cat\'{o}lica del Per\'{u}, Peru;
Ministry of Science and Higher Education and National Science Centre, Poland;
Korea Institute of Science and Technology Information and National Research Foundation of Korea (NRF), Republic of Korea;
Ministry of Education and Scientific Research, Institute of Atomic Physics and Romanian National Agency for Science, Technology and Innovation, Romania;
Joint Institute for Nuclear Research (JINR), Ministry of Education and Science of the Russian Federation, National Research Centre Kurchatov Institute, Russian Science Foundation and Russian Foundation for Basic Research, Russia;
Ministry of Education, Science, Research and Sport of the Slovak Republic, Slovakia;
National Research Foundation of South Africa, South Africa;
Swedish Research Council (VR) and Knut \& Alice Wallenberg Foundation (KAW), Sweden;
European Organization for Nuclear Research, Switzerland;
National Science and Technology Development Agency (NSDTA), Suranaree University of Technology (SUT) and Office of the Higher Education Commission under NRU project of Thailand, Thailand;
Turkish Atomic Energy Agency (TAEK), Turkey;
National Academy of  Sciences of Ukraine, Ukraine;
Science and Technology Facilities Council (STFC), United Kingdom;
National Science Foundation of the United States of America (NSF) and United States Department of Energy, Office of Nuclear Physics (DOE NP), United States of America.
\end{acknowledgement}

\bibliographystyle{utphys}   
\bibliography{biblio}

\newpage
\appendix

%
%

\section{The ALICE Collaboration}
\label{app:collab}

\begingroup
\small
\begin{flushleft}
S.~Acharya\Irefn{org140}\And 
D.~Adamov\'{a}\Irefn{org93}\And 
S.P.~Adhya\Irefn{org140}\And 
A.~Adler\Irefn{org74}\And 
J.~Adolfsson\Irefn{org80}\And 
M.M.~Aggarwal\Irefn{org98}\And 
G.~Aglieri Rinella\Irefn{org34}\And 
M.~Agnello\Irefn{org31}\And 
N.~Agrawal\Irefn{org48}\And 
Z.~Ahammed\Irefn{org140}
% ...
\And
\renewcommand\labelenumi{\textsuperscript{\theenumi}~}

\section*{Affiliation notes}
\renewcommand\theenumi{\roman{enumi}}
\begin{Authlist}
\item \Adef{org*}Deceased
\item \Adef{orgI}Dipartimento DET del Politecnico di Torino, Turin, Italy
\item \Adef{orgII}M.V. Lomonosov Moscow State University, D.V. Skobeltsyn Institute of Nuclear, Physics, Moscow, Russia
\item \Adef{orgIII}Department of Applied Physics, Aligarh Muslim University, Aligarh, India
\item \Adef{orgIV}Institute of Theoretical Physics, University of Wroclaw, Poland
\end{Authlist}

\section*{Collaboration Institutes}
\renewcommand\theenumi{\arabic{enumi}~}
\begin{Authlist}
\item \Idef{org1}A.I. Alikhanyan National Science Laboratory (Yerevan Physics Institute) Foundation, Yerevan, Armenia
\item \Idef{org2}Bogolyubov Institute for Theoretical Physics, National Academy of Sciences of Ukraine, Kiev, Ukraine
\item \Idef{org3}Bose Institute, Department of Physics  and Centre for Astroparticle Physics and Space Science (CAPSS), Kolkata, India
\item \Idef{org4}Budker Institute for Nuclear Physics, Novosibirsk, Russia
\item \Idef{org5}California Polytechnic State University, San Luis Obispo, California, United States
\item \Idef{org6}Central China Normal University, Wuhan, China
\item \Idef{org7}Centre de Calcul de l'IN2P3, Villeurbanne, Lyon, France
\item \Idef{org8}Centro de Aplicaciones Tecnol\'{o}gicas y Desarrollo Nuclear (CEADEN), Havana, Cuba
\item \Idef{org9}Centro de Investigaci\'{o}n y de Estudios Avanzados (CINVESTAV), Mexico City and M\'{e}rida, Mexico
\item \Idef{org10}Centro Fermi - Museo Storico della Fisica e Centro Studi e Ricerche ``Enrico Fermi', Rome, Italy
\item \Idef{org11}Chicago State University, Chicago, Illinois, United States
\item \Idef{org12}China Institute of Atomic Energy, Beijing, China
\item \Idef{org13}Chonbuk National University, Jeonju, Republic of Korea
\item \Idef{org14}Comenius University Bratislava, Faculty of Mathematics, Physics and Informatics, Bratislava, Slovakia
\item \Idef{org15}COMSATS Institute of Information Technology (CIIT), Islamabad, Pakistan
\item \Idef{org16}Creighton University, Omaha, Nebraska, United States
\item \Idef{org17}Department of Physics, Aligarh Muslim University, Aligarh, India
\item \Idef{org18}Department of Physics, Pusan National University, Pusan, Republic of Korea
\item \Idef{org19}Department of Physics, Sejong University, Seoul, Republic of Korea
\item \Idef{org20}Department of Physics, University of California, Berkeley, California, United States
\item \Idef{org21}Department of Physics, University of Oslo, Oslo, Norway
\item \Idef{org22}Department of Physics and Technology, University of Bergen, Bergen, Norway
\item \Idef{org23}Dipartimento di Fisica dell'Universit\`{a} 'La Sapienza' and Sezione INFN, Rome, Italy
\item \Idef{org24}Dipartimento di Fisica dell'Universit\`{a} and Sezione INFN, Cagliari, Italy
\item \Idef{org25}Dipartimento di Fisica dell'Universit\`{a} and Sezione INFN, Trieste, Italy
\item \Idef{org26}Dipartimento di Fisica dell'Universit\`{a} and Sezione INFN, Turin, Italy
\item \Idef{org27}Dipartimento di Fisica e Astronomia dell'Universit\`{a} and Sezione INFN, Bologna, Italy
\item \Idef{org28}Dipartimento di Fisica e Astronomia dell'Universit\`{a} and Sezione INFN, Catania, Italy
\item \Idef{org29}Dipartimento di Fisica e Astronomia dell'Universit\`{a} and Sezione INFN, Padova, Italy
\item \Idef{org30}Dipartimento di Fisica `E.R.~Caianiello' dell'Universit\`{a} and Gruppo Collegato INFN, Salerno, Italy
\item \Idef{org31}Dipartimento DISAT del Politecnico and Sezione INFN, Turin, Italy
\item \Idef{org32}Dipartimento di Scienze e Innovazione Tecnologica dell'Universit\`{a} del Piemonte Orientale and INFN Sezione di Torino, Alessandria, Italy
\item \Idef{org33}Dipartimento Interateneo di Fisica `M.~Merlin' and Sezione INFN, Bari, Italy
\item \Idef{org34}European Organization for Nuclear Research (CERN), Geneva, Switzerland
\item \Idef{org35}Faculty of Electrical Engineering, Mechanical Engineering and Naval Architecture, University of Split, Split, Croatia
\item \Idef{org36}Faculty of Engineering and Science, Western Norway University of Applied Sciences, Bergen, Norway
\item \Idef{org37}Faculty of Nuclear Sciences and Physical Engineering, Czech Technical University in Prague, Prague, Czech Republic
\item \Idef{org38}Faculty of Science, P.J.~\v{S}af\'{a}rik University, Ko\v{s}ice, Slovakia
\item \Idef{org39}Frankfurt Institute for Advanced Studies, Johann Wolfgang Goethe-Universit\"{a}t Frankfurt, Frankfurt, Germany
\item \Idef{org40}Gangneung-Wonju National University, Gangneung, Republic of Korea
\item \Idef{org41}Gauhati University, Department of Physics, Guwahati, India
\item \Idef{org42}Helmholtz-Institut f\"{u}r Strahlen- und Kernphysik, Rheinische Friedrich-Wilhelms-Universit\"{a}t Bonn, Bonn, Germany
\item \Idef{org43}Helsinki Institute of Physics (HIP), Helsinki, Finland
\item \Idef{org44}High Energy Physics Group,  Universidad Aut\'{o}noma de Puebla, Puebla, Mexico
\item \Idef{org45}Hiroshima University, Hiroshima, Japan
\item \Idef{org46}Hochschule Worms, Zentrum  f\"{u}r Technologietransfer und Telekommunikation (ZTT), Worms, Germany
\item \Idef{org47}Horia Hulubei National Institute of Physics and Nuclear Engineering, Bucharest, Romania
\item \Idef{org48}Indian Institute of Technology Bombay (IIT), Mumbai, India
\item \Idef{org49}Indian Institute of Technology Indore, Indore, India
\item \Idef{org50}Indonesian Institute of Sciences, Jakarta, Indonesia
\item \Idef{org51}INFN, Laboratori Nazionali di Frascati, Frascati, Italy
\item \Idef{org52}INFN, Sezione di Bari, Bari, Italy
\item \Idef{org53}INFN, Sezione di Bologna, Bologna, Italy
\item \Idef{org54}INFN, Sezione di Cagliari, Cagliari, Italy
\item \Idef{org55}INFN, Sezione di Catania, Catania, Italy
\item \Idef{org56}INFN, Sezione di Padova, Padova, Italy
\item \Idef{org57}INFN, Sezione di Roma, Rome, Italy
\item \Idef{org58}INFN, Sezione di Torino, Turin, Italy
\item \Idef{org59}INFN, Sezione di Trieste, Trieste, Italy
\item \Idef{org60}Inha University, Incheon, Republic of Korea
\item \Idef{org61}Institut de Physique Nucl\'{e}aire d'Orsay (IPNO), Institut National de Physique Nucl\'{e}aire et de Physique des Particules (IN2P3/CNRS), Universit\'{e} de Paris-Sud, Universit\'{e} Paris-Saclay, Orsay, France
\item \Idef{org62}Institute for Nuclear Research, Academy of Sciences, Moscow, Russia
\item \Idef{org63}Institute for Subatomic Physics, Utrecht University/Nikhef, Utrecht, Netherlands
\item \Idef{org64}Institute for Theoretical and Experimental Physics, Moscow, Russia
\item \Idef{org65}Institute of Experimental Physics, Slovak Academy of Sciences, Ko\v{s}ice, Slovakia
\item \Idef{org66}Institute of Physics, Homi Bhabha National Institute, Bhubaneswar, India
\item \Idef{org67}Institute of Physics of the Czech Academy of Sciences, Prague, Czech Republic
\item \Idef{org68}Institute of Space Science (ISS), Bucharest, Romania
\item \Idef{org69}Institut f\"{u}r Kernphysik, Johann Wolfgang Goethe-Universit\"{a}t Frankfurt, Frankfurt, Germany
\item \Idef{org70}Instituto de Ciencias Nucleares, Universidad Nacional Aut\'{o}noma de M\'{e}xico, Mexico City, Mexico
\item \Idef{org71}Instituto de F\'{i}sica, Universidade Federal do Rio Grande do Sul (UFRGS), Porto Alegre, Brazil
\item \Idef{org72}Instituto de F\'{\i}sica, Universidad Nacional Aut\'{o}noma de M\'{e}xico, Mexico City, Mexico
\item \Idef{org73}iThemba LABS, National Research Foundation, Somerset West, South Africa
\item \Idef{org74}Johann-Wolfgang-Goethe Universit\"{a}t Frankfurt Institut f\"{u}r Informatik, Fachbereich Informatik und Mathematik, Frankfurt, Germany
\item \Idef{org75}Joint Institute for Nuclear Research (JINR), Dubna, Russia
\item \Idef{org76}Korea Institute of Science and Technology Information, Daejeon, Republic of Korea
\item \Idef{org77}KTO Karatay University, Konya, Turkey
\item \Idef{org78}Laboratoire de Physique Subatomique et de Cosmologie, Universit\'{e} Grenoble-Alpes, CNRS-IN2P3, Grenoble, France
\item \Idef{org79}Lawrence Berkeley National Laboratory, Berkeley, California, United States
\item \Idef{org80}Lund University Department of Physics, Division of Particle Physics, Lund, Sweden
\item \Idef{org81}Nagasaki Institute of Applied Science, Nagasaki, Japan
\item \Idef{org82}Nara Women{'}s University (NWU), Nara, Japan
\item \Idef{org83}National and Kapodistrian University of Athens, School of Science, Department of Physics , Athens, Greece
\item \Idef{org84}National Centre for Nuclear Research, Warsaw, Poland
\item \Idef{org85}National Institute of Science Education and Research, Homi Bhabha National Institute, Jatni, India
\item \Idef{org86}National Nuclear Research Center, Baku, Azerbaijan
\item \Idef{org87}National Research Centre Kurchatov Institute, Moscow, Russia
\item \Idef{org88}Niels Bohr Institute, University of Copenhagen, Copenhagen, Denmark
\item \Idef{org89}Nikhef, National institute for subatomic physics, Amsterdam, Netherlands
\item \Idef{org90}NRC Kurchatov Institute IHEP, Protvino, Russia
\item \Idef{org91}NRNU Moscow Engineering Physics Institute, Moscow, Russia
\item \Idef{org92}Nuclear Physics Group, STFC Daresbury Laboratory, Daresbury, United Kingdom
\item \Idef{org93}Nuclear Physics Institute of the Czech Academy of Sciences, \v{R}e\v{z} u Prahy, Czech Republic
\item \Idef{org94}Oak Ridge National Laboratory, Oak Ridge, Tennessee, United States
\item \Idef{org95}Ohio State University, Columbus, Ohio, United States
\item \Idef{org96}Petersburg Nuclear Physics Institute, Gatchina, Russia
\item \Idef{org97}Physics department, Faculty of science, University of Zagreb, Zagreb, Croatia
\item \Idef{org98}Physics Department, Panjab University, Chandigarh, India
\item \Idef{org99}Physics Department, University of Jammu, Jammu, India
\item \Idef{org100}Physics Department, University of Rajasthan, Jaipur, India
\item \Idef{org101}Physikalisches Institut, Eberhard-Karls-Universit\"{a}t T\"{u}bingen, T\"{u}bingen, Germany
\item \Idef{org102}Physikalisches Institut, Ruprecht-Karls-Universit\"{a}t Heidelberg, Heidelberg, Germany
\item \Idef{org103}Physik Department, Technische Universit\"{a}t M\"{u}nchen, Munich, Germany
\item \Idef{org104}Research Division and ExtreMe Matter Institute EMMI, GSI Helmholtzzentrum f\"ur Schwerionenforschung GmbH, Darmstadt, Germany
\item \Idef{org105}Rudjer Bo\v{s}kovi\'{c} Institute, Zagreb, Croatia
\item \Idef{org106}Russian Federal Nuclear Center (VNIIEF), Sarov, Russia
\item \Idef{org107}Saha Institute of Nuclear Physics, Homi Bhabha National Institute, Kolkata, India
\item \Idef{org108}School of Physics and Astronomy, University of Birmingham, Birmingham, United Kingdom
\item \Idef{org109}Secci\'{o}n F\'{\i}sica, Departamento de Ciencias, Pontificia Universidad Cat\'{o}lica del Per\'{u}, Lima, Peru
\item \Idef{org110}Shanghai Institute of Applied Physics, Shanghai, China
\item \Idef{org111}St. Petersburg State University, St. Petersburg, Russia
\item \Idef{org112}Stefan Meyer Institut f\"{u}r Subatomare Physik (SMI), Vienna, Austria
\item \Idef{org113}SUBATECH, IMT Atlantique, Universit\'{e} de Nantes, CNRS-IN2P3, Nantes, France
\item \Idef{org114}Suranaree University of Technology, Nakhon Ratchasima, Thailand
\item \Idef{org115}Technical University of Ko\v{s}ice, Ko\v{s}ice, Slovakia
\item \Idef{org116}Technische Universit\"{a}t M\"{u}nchen, Excellence Cluster 'Universe', Munich, Germany
\item \Idef{org117}The Henryk Niewodniczanski Institute of Nuclear Physics, Polish Academy of Sciences, Cracow, Poland
\item \Idef{org118}The University of Texas at Austin, Austin, Texas, United States
\item \Idef{org119}Universidad Aut\'{o}noma de Sinaloa, Culiac\'{a}n, Mexico
\item \Idef{org120}Universidade de S\~{a}o Paulo (USP), S\~{a}o Paulo, Brazil
\item \Idef{org121}Universidade Estadual de Campinas (UNICAMP), Campinas, Brazil
\item \Idef{org122}Universidade Federal do ABC, Santo Andre, Brazil
\item \Idef{org123}University College of Southeast Norway, Tonsberg, Norway
\item \Idef{org124}University of Cape Town, Cape Town, South Africa
\item \Idef{org125}University of Houston, Houston, Texas, United States
\item \Idef{org126}University of Jyv\"{a}skyl\"{a}, Jyv\"{a}skyl\"{a}, Finland
\item \Idef{org127}University of Liverpool, Liverpool, United Kingdom
\item \Idef{org128}University of Science and Techonology of China, Hefei, China
\item \Idef{org129}University of Tennessee, Knoxville, Tennessee, United States
\item \Idef{org130}University of the Witwatersrand, Johannesburg, South Africa
\item \Idef{org131}University of Tokyo, Tokyo, Japan
\item \Idef{org132}University of Tsukuba, Tsukuba, Japan
\item \Idef{org133}Universit\'{e} Clermont Auvergne, CNRS/IN2P3, LPC, Clermont-Ferrand, France
\item \Idef{org134}Universit\'{e} de Lyon, Universit\'{e} Lyon 1, CNRS/IN2P3, IPN-Lyon, Villeurbanne, Lyon, France
\item \Idef{org135}Universit\'{e} de Strasbourg, CNRS, IPHC UMR 7178, F-67000 Strasbourg, France, Strasbourg, France
\item \Idef{org136} Universit\'{e} Paris-Saclay Centre d'\'Etudes de Saclay (CEA), IRFU, Department de Physique Nucl\'{e}aire (DPhN), Saclay, France
\item \Idef{org137}Universit\`{a} degli Studi di Foggia, Foggia, Italy
\item \Idef{org138}Universit\`{a} degli Studi di Pavia, Pavia, Italy
\item \Idef{org139}Universit\`{a} di Brescia, Brescia, Italy
\item \Idef{org140}Variable Energy Cyclotron Centre, Homi Bhabha National Institute, Kolkata, India
\item \Idef{org141}Warsaw University of Technology, Warsaw, Poland
\item \Idef{org142}Wayne State University, Detroit, Michigan, United States
\item \Idef{org143}Westf\"{a}lische Wilhelms-Universit\"{a}t M\"{u}nster, Institut f\"{u}r Kernphysik, M\"{u}nster, Germany
\item \Idef{org144}Wigner Research Centre for Physics, Hungarian Academy of Sciences, Budapest, Hungary
\item \Idef{org145}Yale University, New Haven, Connecticut, United States
\item \Idef{org146}Yonsei University, Seoul, Republic of Korea
\end{Authlist}
\endgroup
  
\end{document}